\documentclass[letter]{jpsj3}
\usepackage{txfonts}
\usepackage{color}

\begin{document}

\title{Critical Crossover Between Yosida-Kondo Dominant Regime and Magnetic Frustration Dominant Regime in the System of a Magnetic Trimer on a Metal Surface}

\author{Nghiem Thi Minh Hoa\thanks{E-mail address: hoa@dyn.ap.eng.osaka-u.ac.jp}, Wilson Agerico Di{\~n}o, and Hideaki Kasai}

\inst{Department of Applied Physics, Osaka University, Suita, Osaka 565-0871, Japan\\\quad\\\bf(revised on \today)}

\abst{ Quantum Monte Carlo simulations were carried out for the system of a magnetic trimer on a metal surface. The magnetic trimer is arranged in two geometric configurations, viz., isosceles and equilateral triangles. The calculated spectral density and magnetic susceptibility show the existence of two regimes: Yosida-Kondo dominant regime and magnetic frustration dominant regime. Furthermore, a critical crossover between these two regimes can be induced by changing the configuration of the magnetic trimers from isosceles to equilateral triangle.}

\kword{magnetic trimers on surface, magnetic frustration, Kondo effect, Yosida-Kondo resonance, critical crossover.}

\maketitle

Studies on systems of magnetic monomers and dimers on a metal surface have allowed us to gain deeper understanding of, and also clarify, the nature of typical phenomena associated with the classical Kondo systems, e.g., the real-space image of Yosida-Kondo resonance and the competition between the spin-flip interaction and the indirect Ruderman-Kittel-Kasuya-Yosida (RKKY) interaction\cite{kawa,kasai,mad,Emi,Emi2,Ott}. Furthermore, in the system of a magnetic trimer on a metal surface \cite{Hoa}, we found that changing the geometric configuration of the trimer, from isosceles to equilateral triangle, changes the corresponding spectral density near the Fermi level. In the isosceles configuration, the spectral density exhibits a sharp peak near the Fermi level, which we attribute to the Yosida-Kondo resonance, with a corresponding relatively high Kondo temperature. While in the equilateral configuration, no peak is observed near the Fermi level\cite{Hoa}. These results are in agreement with scanning tunneling spectroscopy (STS) experiments on Cr trimer on Au(111)\cite{Jam}. These results also indicate the existence of two separate regimes. One is a Yosida-Kondo dominant regime for an isosceles trimer where the ground-state of the system is singlet. The other regime is a magnetic frustration dominant regime for an equilateral trimer, in which the ground-state of the system is degenerate due to the competing antiferromagnetic (AF) interactions between adatom pairs. 

\begin{figure}[ht]\begin{center}
\includegraphics[scale=0.5]{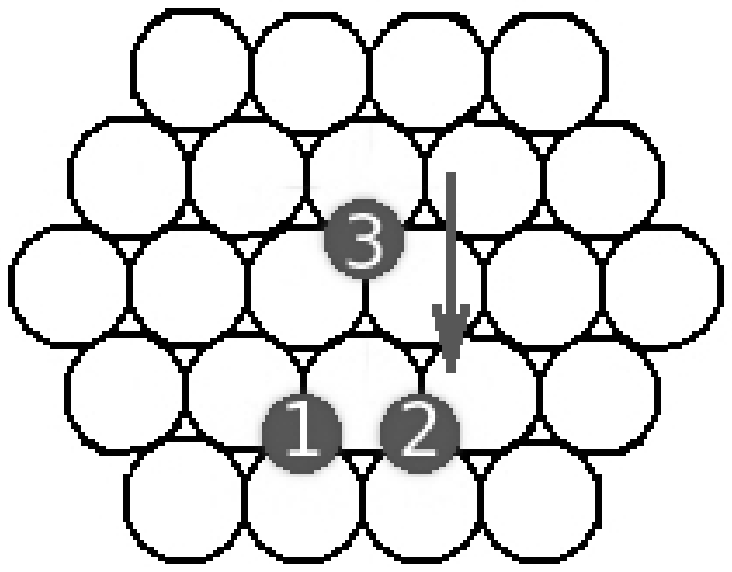}
\caption{ Magnetic trimer on a metal surface. Separation between adatoms 1 and 2, $R_{12}=a$. $a$ is the surface lattice constant. Adatom 3 is moved along the direction of the arrow, maintaining the relation $R_{13}=R_{23}$. \label{fig:3atom}}\end{center}
\end{figure}

Can we observe the crossover between the Yosida-Kondo dominant regime and the magnetic frustration dominant regime? This is the main concern in this paper. To answer this question, we again consider the system of magnetic trimer on a metal surface, as shown in Fig. \ref{fig:3atom}. By calculating the spectral density, as well as the magnetic susceptibility at adatom 3 for a wide range of temperature, we clarify the existence of two separate regimes and suggest the nature of the crossover between these two regimes.

A system of three magnetic adatoms on a metal surface is modeled as follows,

\begin{eqnarray}
\label{H}
H=\sum_i\epsilon_{i\sigma}n_{i\sigma}+\sum_i U_in_{i\uparrow}n_{i\downarrow}
+\sum_{\sigma,\langle ij\rangle} (t_{ij}d^+_{i\sigma}d_{j\sigma}\nonumber+h.c)\\
+\sum_{\mathbf{k}\sigma} \epsilon_{\mathbf{k}} n_{\mathbf{k}\sigma}+\sum_{\mathbf{k}\sigma,i} (V_{\mathbf{k}i}c^+_{\mathbf{k}\sigma}d_{i\sigma}+h.c).
\end{eqnarray}
Adatom indices $i,j=1,2,3$. Spin index $\sigma=\uparrow,\downarrow$. $\epsilon_{i}$ is the energy level associated with the electrons at adatom site $i$. $U_i$ gives the Coulomb repulsion at individual sites. $t_{ij}$ is the direct hopping between the adatoms ($i \neq j$), and gives rise to the direct AF interactions. $\mathbf{k}$ is the metal surface conduction electron wave-vector. $V_{\mathbf{k}i}=e^{-i\mathbf{k}\mathbf{R}_i}V_0$ gives the hybridization between the states localized at the adatoms and the conduction electron medium. $\mathbf{R}_i$ is the adatom position. We choose a free-electron dispersion relation for the 2D conduction electron medium, i.e.,

\begin{equation}\label{disper}\epsilon_{\mathbf{k}}=\frac{D}{2}[(\frac{k}{k_{\mathrm{F}}})^2-1].\end{equation}
$D$ is the conduction electron bandwidth and $k_{\mathrm{F}}$ is the Fermi wave-number. Since the wave-number is within the range $0<k<\sqrt{2}k_{\mathrm{F}}$, then $-\frac{D}{2}<\epsilon_k<\frac{D}{2}$.

The Green's functions are defined so as to satisfy the equation

\begin{equation}
\label{eom}
\sum_{\nu}(i\omega_n-H)_{\mu\nu}G_{\nu\kappa}(i\omega_n)=\delta_{\mu\kappa},
\end{equation}
with adatom indices $i,j$ and wave-vector of conduction electrons $\mathbf{k}$ given by $\mu$, $\nu$, and $\kappa$. $\omega_n=(2n+1)\pi/\beta$. $n$ is an integral number. $\beta=1/k_{\mathrm{B}}T$ (eV)$^{-1}$. $k_{\mathrm{B}}$ is the Boltzmann constant. From eqn. \eqref{eom}, we derive the non-interacting Green's function, which corresponds to $U_i=0$ case, and get

\begin{equation}
\label{green0} G_{ij}^0(i\omega_n)=\left( 
\begin{array}{ccc}
i\omega_n-\tilde{\epsilon}_{1}-F_0 & -t_{12}-F_{12} & -t_{13}-F_{13} \\
-t_{21}-F_{21} & i\omega_n-\tilde{\epsilon}_{2}-F_0& -t_{23}-F_{23} \\
-t_{31}-F_{31} & -t_{32}-F_{32} & i\omega_n-\tilde{\epsilon}_{3}-F_0 \end{array} \right)^{-1}
\end{equation}
with $\tilde{\epsilon}_{i}={\epsilon}_{i}+U_i/2$. $F_0$ and $F_{ij}$ ($i\neq j$) are as given in our previous study\cite{Hoa2}. 

\begin{equation}
\label{F0}
F_{0}=\sum_{\mathbf{k}} \frac{V_{0}^2}{i\omega_n-\epsilon_{\mathbf{k}}}=\frac{\Delta D}{\pi}\int_0^{\sqrt{2}}\frac{xdx}{i\omega_n-\frac{D}{2}(x^2-1)}
\end{equation}

\begin{equation}
\label{F12}
F_{ij}=\sum_{\mathbf{k}} \frac{V_{0}^2 e^{-i\mathbf{k}\mathbf{R}_{ij}}}{i\omega_n-\epsilon_{\mathbf{k}}}=\frac{\Delta D}{\pi}\int_0^{\sqrt{2}}\frac{xJ_0(xk_{\mathrm{F}}R_{ij})dx}{i\omega_n-\frac{D}{2}(x^2-1)}
\end{equation}
$\mathbf{R}_{ij}=\mathbf{R}_i-\mathbf{R}_j$, $\Delta=V_0^2k_{\mathrm{F}}^2/2D$, and $J_0(xk_{\mathrm{F}}R_{12})$ is the $0^{th}$ order Bessel function of the first kind. $F_0(i\omega_n/\Delta)$ accounts for the finite peak width at $\epsilon_{i}$, $\epsilon_{i}+U_i$, and also the Yosida-Kondo resonance at the Fermi level, at low temperatures. $F_{ij}(i\omega_n/\Delta,k_{\mathrm{F}}R_{ij})$ gives rise to the RKKY interaction between adatoms.

In this study, we used quantum Monte Carlo (QMC) simulations to get the interacting Green's function corresponding to the $U_i\neq0$ case and the magnetic susceptibility at each adatom\cite{HirschFye,HirschFye2,note1}. We used the maximum entropy method to get the spectral densities from the Green's functions\cite{Silver}. 

\begin{figure}[ht]\begin{center}
\includegraphics[scale=0.678]{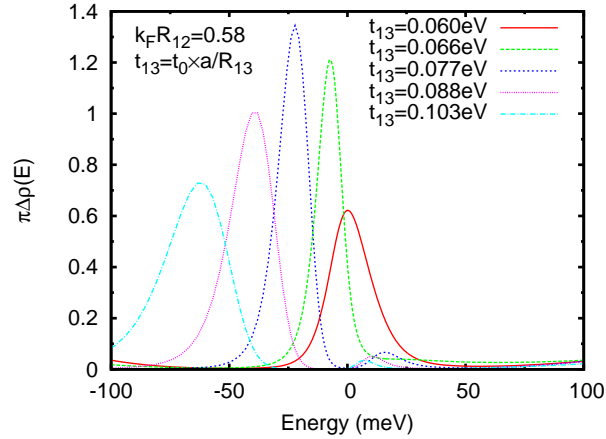}
\caption{(Color online) Spectral densities at adatom 3 in the system of Cr trimer on Au(111) at 54K. The Fermi level is set as the energy reference. Parameters corresponding to Au(111) are set as follows: lattice constant $a=2.9$\AA, Fermi wave-number $k_{\mathrm{F}}=0.2$\AA$^{-1}$, and the half-bandwidth $D/2=0.5$eV\cite{kevan}. For Cr adatoms, the energy level of $d$ electrons $\epsilon_d=-0.15$eV, Coulomb repulsion $U=0.3$eV, and 
$t_{12}=0.103$eV. $\Delta=0.015$eV results from the coupling between adatoms and surface and gives the broadening of $d$ levels\cite{Hoa}. As $k_{\mathrm{F}}R_{12}=0.58$ and $k_{\mathrm{F}}R_{13}<1.6$ then the RKKY interaction between adatoms rising from $F_{ij}$ in eqn. \eqref{green0} is  ferromagnetic\cite{Hoa2}. In the system specified above, the magnitude of the AF interaction rising from $t_{ij}$ is much larger than that of the RKKY interaction, then the calculated densities are shown as $t_{13}$ dependence.\label{fig:decreaseAu}}\end{center}
\end{figure}

From hereon, we consider the specific case of a Cr trimer on Au(111), in the configuration depicted in Fig. \ref{fig:3atom}. The separation between adatoms 1 and 2 is kept constant, while adatom 3 is shifted. We assume that the direct hopping between adatoms $t_{ij}$ is inversely proportional to adatom separation $R_{ij}$. Since adatoms 1 and 2 are fixed, we can set $t_{12}$ equal to a constant. By further setting $t_{13}=t_{23}=t_{12}\times R_{12}/R_{13}$, the direct AF interaction rising from $t_{13}$ can be tuned by varying $R_{13}$. The corresponding spectral densities at adatom 3 are shown in Fig. \ref{fig:decreaseAu}.

When  $R_{13}=R_{23}\gg R_{12}$, adatom 3 on Au(111) forms a single impurity Kondo system with a corresponding extremely low Kondo temperature. At high temperatures, the magnetic moment of adatom 3 is not fully compensated by the conduction electrons, therefore no Yosida-Kondo resonance would be observed\cite{Hoa,Jam}. The dimer, which consists of adatoms 1 and 2, is in the singlet state $\{\uparrow\downarrow\}$ due to the strong AF interaction. 

When adatom 3 is moved closer to the dimer, we found that the resonance at the Fermi level is most pronounced for the isosceles trimer with $R_{13}=5.0$\AA, corresponding to $t_{13}=0.06$eV (the red/solid curve in Fig. \ref{fig:decreaseAu}). We expect that the dominance of the Kondo effect at adatom 3 is enhanced by the combination of AF interactions between adatoms, since adatom 3 is dependent not only on the magnetic interactions with other adatoms, but also on the interaction between adatoms 1 and 2\cite{Hoa}. The ground state of this isosceles trimer system is $\{\uparrow\downarrow\bullet\}$, where $\{\bullet\}$ stands for the Yosida-Kondo singlet. 

When $R_{13}=R_{12}$ corresponding to $t_{13}=0.103$eV,  the high peak structure at the Fermi level disappears (the blue/dot-dashed curve in Fig. \ref{fig:decreaseAu}). The competition of the equally strong AF interactions between the adatom pairs gives rise to the dominance of magnetic frustration. In terms of the spin model, $H=\sum_{<i,j>}JS_iS_j$ with $J>0$ and $S=\pm 1/2$, the ground state of the equilateral trimer can either be $\{\uparrow\downarrow\uparrow\}$ or $\{\uparrow\downarrow\downarrow\}$, associated with the energy $E=-J/4$ lower than the Fermi level, while the excited state $\{\uparrow\uparrow\uparrow\}$ associates with the energy $E=3J/4$. Therefore, we observe the Yosida-Kondo resonance suppressed due to the dominance of magnetic frustration in the equilateral trimer.


In Fig. \ref{fig:decreaseAu}, as $t_{13}$ increases from $0.06$eV, the spectral density at the Fermi level drastically reduces. We observe the resonance feature shifting away from the Fermi level and merging with the $d$ resonance. The small peak rising near the Fermi level is the Yosida-Kondo resonance. This behavior is different from that observed in the dimer system, where the Yosida-Kondo resonance splits into two nearly symmetric (with respect to the Fermi level) peaks\cite{Emi,Ott}. Thus, we expect a difference in the nature of the crossover behavior of a trimer system and a dimer system. In the dimer system, the crossover is from the Yosida-Kondo singlet state at each adatom to the AF singlet state $\{\uparrow\downarrow\}$. In the trimer system, the crossover at adatom 3 is from the Yosida-Kondo singlet state to the degenerate states as a result of magnetic frustration, in which spin direction can either be $\{\uparrow\}$ or $\{\downarrow\}$. 

\begin{figure}[t]\begin{center}
\includegraphics[scale=0.94]{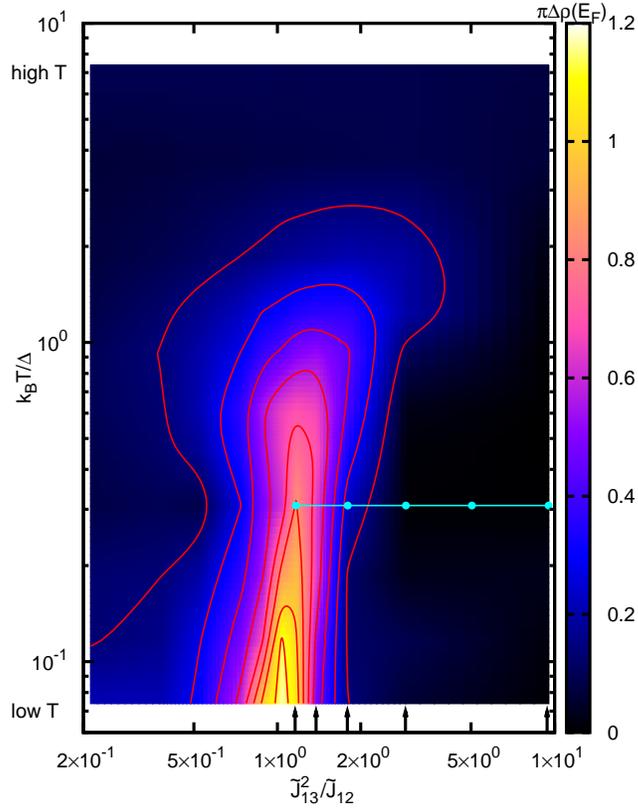}
\caption{(Color online) Contour plot of the spectral densities at the Fermi level at adatom 3 versus temperature and magnetic interaction, the contour spacing is 0.14. At each blue (white) point, the spectral density as a function of energy is presented in Fig. \ref{fig:decreaseAu}.\label{fig:compare}}\end{center}
\end{figure}

In order to track the two separate regimes and observe the crossover between them, we study the temperature dependence of the spectral density at the Fermi level. Using the standard definition of AF interaction rising from the direct hopping $t$ and scaling with $\Delta$\cite{YoYa}, we have $\tilde{J}_{12}=4t_{12}^2/U\Delta$, and $\tilde{J}_{13}=4t_{13}^2/U\Delta$. From this point on, we use $\tilde{J}_{13}^2/\tilde{J}_{12}$ as a variable, instead of $t_{13}$, in order to present the dependence of properties at adatom 3 not only on $\tilde{J}_{13}$, but also on the interaction between adatom 1 and 2. The equilateral trimer corresponds to $\tilde{J}_{13}^2/\tilde{J}_{12}=\tilde{J}_{12}\approx 10$, while $\tilde{J}_{13}^2/\tilde{J}_{12}\ll 1$ corresponds to the case when adatom 3 is sufficiently far from adatoms 1 and 2, and forms a single impurity Kondo system with $k_{\mathrm{B}}T_{\mathrm{K}}^{single}/\Delta=\pi\times10^{-3}$. Within the intermediate range $\tilde{J}_{13}^2/\tilde{J}_{12}$, we show the spectral density at the Fermi level in a broad range of $k_{\mathrm{B}}T/\Delta$ in Fig. \ref{fig:compare}. The spectral densities at the Fermi level for various $t_{13}$ shown in Fig. \ref{fig:decreaseAu} correspond to the blue (white) dots in Fig. \ref{fig:compare} at $k_{\mathrm{B}}T/\Delta\approx0.3$.

When $\tilde{J}_{13}^2/\tilde{J}_{12}\ge2$, the corresponding density at the Fermi level is almost zero, not only at a finite temperature $k_{\mathrm{B}}T/\Delta\approx0.3$, but also in a broad range of temperatures. 
In the range $2>\tilde{J}_{13}^2/\tilde{J}_{12}>1$, the lower the temperature, the more significant the increase in density with decreasing $\tilde{J}_{13}^2/\tilde{J}_{12}$. 
When $\tilde{J}_{13}^2/\tilde{J}_{12}\le1$, the density decreases more gradually at lower temperatures. Thus, we observe two separate regimes, viz., the magnetic frustration dominant regime in $\tilde{J}_{13}^2/\tilde{J}_{12}\ge2$ where the density at the Fermi level is nearly equal to zero; and the Yosida-Kondo dominant regime in the range $\tilde{J}_{13}^2/\tilde{J}_{12}\le1$ where the density at the Fermi level increases as the temperature decreases. The crossover point between two regimes should appear in the range  $2>\tilde{J}_{13}^2/\tilde{J}_{12}>1$ where the variation of density versus $\tilde{J}_{13}^2/\tilde{J}_{12}$ is drastic at low temperatures.

\begin{figure}[t]
\begin{center}
\includegraphics[scale=0.678]{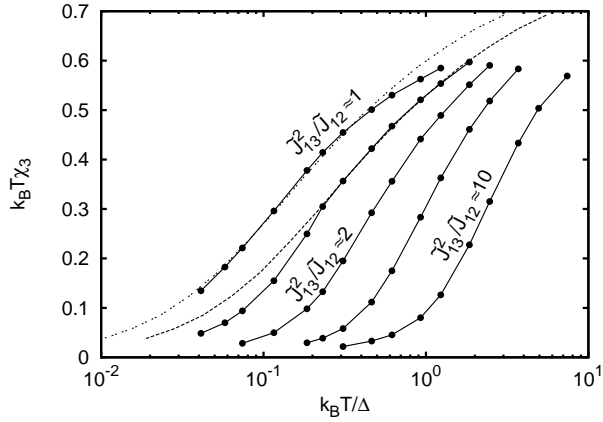}
\caption{ Temperature dependence of the magnetic susceptibilities at adatom 3 (solid lines) at magnitudes of interactions corresponding to the arrows indicated in Fig. \ref{fig:compare}. Dashed lines depict the universal shape of Kondo susceptibility with the effective Kondo temperatures $k_{\mathrm{B}}T_{\mathrm{K}}^*/\Delta=0.1$ and $0.2$\cite{kww,kww2}. The error at each calculated point is smaller than or equal to the radius of the solid circle in the figure.\label{fig:sus}}
\end{center}
\end{figure}

To further clarify the nature of the crossover between the Yosida-Kondo dominant regime and the magnetic frustration regime, we study the temperature dependence of magnetic susceptibility at adatom 3. The magnetic susceptibility at adatom 3 is defined as $\chi_3=\int_0^\beta d\tau\langle\sigma_3(\tau)\sigma_3(0)\rangle$\cite{HirschFye2}. $\sigma_i=n_i^{\uparrow}-n_i^{\downarrow}$. Fig. \ref{fig:sus} shows $k_{\mathrm{B}}T\chi_3$ versus $k_{\mathrm{B}}T/\Delta$ for various $\tilde{J}_{13}^2/\tilde{J}_{12}$, corresponding to the arrows in Fig. \ref{fig:compare}, with the corresponding order from right to left. These plots exhibit three different behaviors, indicating the existence of two separate regimes and the crossover between them.

$\tilde{J}_{13}^2/\tilde{J}_{12}\ge2$ corresponds to the magnetic frustration dominant regime. The corresponding curves of $k_{\mathrm{B}}T\chi_3$ with decreasing $k_{\mathrm{B}}T/\Delta$ do not follow the universal curve of Kondo susceptibility, but decrease rapidly and smoothly. 

$\tilde{J}_{13}^2/\tilde{J}_{12}\approx1$ belongs to the Yosida-Kondo dominant regime. The corresponding curve of $k_{\mathrm{B}}T\chi_3$ versus $k_{\mathrm{B}}T/\Delta$ follows the universal curve of Kondo susceptibility for a single impurity case with the effective Kondo temperature $k_{\mathrm{B}}T_{\mathrm{K}}^*/\Delta=0.1$\cite{kww,kww2}.

$k_{\mathrm{B}}T\chi_3$ versus $k_{\mathrm{B}}T/\Delta$, corresponding to $2>\tilde{J}_{13}^2/\tilde{J}_{12}>1$, also follows the universal curve with $k_{\mathrm{B}}T_{\mathrm{K}} ^*/\Delta=0.2$ at high temperatures. However, as $T<T_{\mathrm{K}}^*$, $k_{\mathrm{B}}T\chi_3$ versus $k_{\mathrm{B}}T/\Delta$ does not continue following the universal curve, but instant drops rapidly close to zero. 

In summary, as we move adatom 3 towards the AF quenched singlet of adatoms 1 and 2 so as $\tilde{J}_{13}^2/\tilde{J}_{12}\le1$, adatom 3 coupling to the metal surface effectively behaves as a single impurity Kondo system but with a higher Kondo temperature than that corresponding to $\tilde{J}_{13}^2/\tilde{J}_{12}\ll 1$. In the particular case of $\tilde{J}_{13}^2/\tilde{J}_{12}\approx1$, $k_{\mathrm{B}}T_{\mathrm{K}}^*/\Delta=0.1\gg k_{\mathrm{B}}T_{\mathrm{K}}^{single}/\Delta=\pi \times 10^{-3}$. And as expected, the crossover is found to occur in the range $1<\tilde{J}_{13}^2/\tilde{J}_{12}<2$ via the study on the temperature dependence of $k_{\mathrm{B}}T\chi_3$. At a finite value of $\tilde{J}_{13}^2/\tilde{J}_{12}$, we observe the discontinuous plot of $k_{\mathrm{B}}T\chi_3$ versus $k_{\mathrm{B}}T/\Delta$. Therefore, we suggest that the critical crossover would be observed in a system of a trimer on a metal surface. As $10\ge\tilde{J}_{13}^2/\tilde{J}_{12}\ge2$, no Kondo temperature is defined due to the dominance of the magnetic frustration.

In conclusion, we found the existence of two separate regimes in a system of a magnetic trimer on a metal surface: the Yosida-Kondo dominant regime and the magnetic frustration dominant regime. Moreover, we suggest the critical crossover between these two above regimes.

\section*{Acknowledgements}
N. T. M. Hoa is supported by MEXT (Ministry of Education, Culture, Sports, Science and Technology) through the Osaka University Quantum Engineering Design course (QEDC) and through the G-COE (Special Coordination Funds for the Global Center of Excellence) program "Atomically Controlled Fabrication Technology". The authors would like to thank E. Minamitani, RIKEN, for the valuable discussions. Some of the numerical calculations presented here were performed at the computer facilities of the Cyber Media Center (Osaka University), YITP (Kyoto University), and T2K (Tsukuba University)


\begin{thebibliography}{99}
\bibitem{note1}In the Hirsch-Fye algorithm for QMC simulation, the spacing between imaginary-times, $\delta\tau=\beta/L$ in which $L$ is a number of time steps, must be kept sufficiently small to minimize systematic Trotter breakup errors\cite{Silver}. In the calculations for the system presented here, the parameters are chosen such that $\delta\tau^2\Delta U<0.015$.
\end{thebibliography}
\end{document}